\documentclass[12pt,article]{aastex}

\usepackage{graphicx,natbib,textcomp,caption,subcaption}         
\usepackage[utf8]{inputenc}
\usepackage{xcolor,amsmath}
\usepackage{multirow}
\usepackage{float}

\slugcomment{ApJ manuscript, \today}

\shorttitle{Nature of Els\"{a}sser variables}
\shortauthors{Magyar, Van Doorsselaere, \& Goossens}

\begin{document}

\title{The nature of Els\"{a}sser variables in compressible MHD}

\author{N. Magyar, T. 
Van Doorsselaere, M. Goossens}
\affil{Centre for mathematical Plasma Astrophysics (CmPA), KU 
Leuven, Celestijnenlaan 
200B bus 2400, 3001 Leuven, Belgium; norbert.magyar@kuleuven.be}

\begin{abstract}

The Els\"{a}sser variables are often used in studies of plasma turbulence, in helping differentiate between MHD waves propagating parallel or anti-parallel to the main magnetic field. While for pure Alfv\'en waves in a  homogeneous  plasma the method is strictly valid, we show that  compressible, magnetoacoustic waves  are in general described by both Els\"{a}sser variables.  Furthermore, in a compressible and inhomogeneous plasma, the pure MHD waves (Alfv\'en, fast and slow) are no longer normal modes, but waves become linearly coupled or display mixed properties of Alfv\'en and magnetoacoustic nature. These waves are necessarily described by both Els\"{a}sser variables and therefore the  Els\"{a}sser  formalism cannot be used to strictly separate parallel and anti-parallel propagating waves. Nevertheless, even in an inhomogeneous plasma, for a highly Alfv\'enic wave the Els\"{a}sser variable corresponding to the propagation direction appears still dominating. We suggest that for Alfv\'enic waves, the relative amplitude of Els\"{a}sser variables depends on the local degree of inhomogeneity and other plasma and wave properties.  This finding has implications for turbulence studies in inhomogeneous and compressible plasmas, such as the solar corona and solar wind.

\end{abstract}


\section{Introduction}

In a short letter, \citet{1950PhRv...79..183E} showed that transforming the incompressible MHD equations, by using the variables now named after him, leads to a symmetrical form of the equations. These variables, as will be shown in the next section, represent pure Alfv\'en wave perturbations propagating either in the direction of the background magnetic field $\mathbf{B}_0$ or opposite to it. This simple dichotomy of wavelike perturbations in plasmas by using the Els\"{a}sser variables turned out to be very useful in studying plasma turbulence, e.g. in the solar wind \citep{2013LRSP...10....2B}, both theoretically \citep[e.g.][]{1980A&A....83...26D,1989JPlPh..41..479M,1989GeoRL..16..755Z}, and for in-situ data analysis \citep[e.g.][]{1989JGR....9411739T,1990JGR....95.8197G}. This was made possible by assuming incompressibility, due to the usually highly Alfv\'enic nature of solar wind perturbations \citep[especially the fast solar wind,][]{2013LRSP...10....2B}. However, it is well known that  the solar wind is compressible and inhomogeneous (especially the slow solar wind), and the nature and origin of the Els\"{a}sser component corresponding to `inward' propagation (propagating towards the Sun) is still not completely clear. These may very well represent locally generated, inward propagating Alfv\'enic waves \citep{1989JGR....9411977B,1989JGR....9411739T}, however this interpretation was found unlikely in a number of studies, suggesting instead that they are either signatures of convected background structures \citep[e.g. pressure-balanced structures in the solar wind,][]{1991JGR....96.7841B,1992JGR....9719129B,1995SSRv...73....1T} or of the compressive component of the perturbations \citep{1993JGR....9821045M,1996AIPC..382..229B}. For example, \citet{1990JGR....95.8197G} found that amplitudes of inward propagating modes are correlated with plasma density perturbations.  The presence of compressibility  introduces new wave modes, i.e. magnetoacoustic modes , making the study of compressible MHD turbulence much more difficult, about which little is known, compared to its incompressible counterpart \citep{0004-637X-562-1-279,PhysRevLett.88.245001}. \citet{1987JGR....92.7363M} showed that the compressible MHD euqations (with a polytropic equation of state) can still be written in terms of generalized Els\"{a}sser variables, with variable density. In the paper, it was also suggested that these equations might be suitable to describe compressible MHD turbulence. Indeed, Els\"{a}sser variables were used regularly in studies of, e.g. solar wind turbulence, even when  inhomogeneities or  density perturbations were present. Moreover, the meaning of Els\"{a}sser variables, i.e.  the general separation of waves into outward and inward propagating components , was extended unchanged to compressible scenarios from the originally incompressible  and homogeneous  framework.  However, even in an incompressible plasma which is inhomogeneous along the magnetic field, the outward and inward propagating Alfv\'en waves are linearly coupled (reflection), resulting in outward propagating waves necessarily described by a `principal' and `anomalous' component when expressed using Els\"{a}sser variables \citep{1973JGR....78.3643H,1980JGR....85.1311H}. Therefore, when reflections occur, even Alfv\'en waves are described by both Els\"{a}sser fields propagating in the same direction, thus they cannot be strictly separated in inward and outward propagating contributions by using the Els\"{a}sser variables \citep{1990JGR....9514873H}. \par 
 In this paper, we show that the presence of compressibility and inhomogeneities across the magnetic field might further aggravate the inability of Els\"{a}sser variables to separate perturbations into inward and outward propagating modes. Magnetoacoustic waves, which display compression, are in general described by both Els\"{a}sser variables, even in a homogeneous plasma. Furthermore, the presence of both compressibility and plasma inhomogeneity allows for the linear coupling of magnetoacoustic and Alfv\'en waves.  Therefore, in a generally inhomogeneous medium, waves are not in their `pure' state: one cannot decompose them into pure fast, slow, and Alfv\'en components, as the waves have mixed properties \citep{2011SSRv..158..289G}. Waves with mixed properties are described by both Els\"{a}sser variables as they propagate, to a varying degree depending  among other factors  on the local plasma inhomogeneity. Previously, this property was used to explain the transition to a turbulent state of an inhomogeneous plasma perturbed by unidirectionally propagating Alfv\'enic waves \citep{2017NatSR...14820M}.  We would like to point out that the adjective ‘Alfvénic’ describes waves which have largely Alfvén characteristics, however, due to plasma inhomogeneity they are not pure Alfvén waves, as compression is also present. Alfvénic waves are an example of MHD waves with mixed properties. \citep{2009A&A...503..213G,2012ApJ...753..111G} . In the following, in Section~\ref{two} we present some simple analytical calculations of the Els\"{a}sser variables for magnetoacoustic modes in infinite and homogeneous plasma. In Section~\ref{three} we present the results of a 2.5D MHD simulation, a `toy model' used to demonstrate the analytically derived results of section~\ref{two}. In Section~\ref{four}, including the effects of inhomogeneity, we consider a 3D toy model to demonstrate the linear coupling of MHD waves and their appearance using the Els\"{a}sser formalism. Finally, in Section~\ref{five}, we conclude the presented results. 

\section{Mathematical formulation}\label{two}

For studying the nature of Els\"{a}sser variables in a compressible plasma, we use the ideal MHD equations \citep{2004prma.book.....G}:
\begin{align}
 \label{continuity}
 &\frac{\partial \rho}{\partial t} + \nabla \cdot (\rho \mathbf{v}) = 0, \\
 \label{motion}
 &\rho \frac{\partial \mathbf{v}}{\partial t} + \rho \mathbf{v} \cdot \nabla \mathbf{v} = -\nabla p + \mathbf{j} \times \mathbf{B}, \\
 \label{energy}
 &\frac{\partial p}{\partial t} + \mathbf{v} \cdot \nabla p + \gamma p \nabla \cdot \mathbf{v} = 0, \\
 \label{induction}
 &\frac{\partial \mathbf{B}}{\partial t} = \nabla \times (\mathbf{v} \times \mathbf{B}), \\
 \label{divb}
 &\nabla \cdot \mathbf{B} = 0,
\end{align}
where $\mathbf{j} = \frac{1}{\mu}(\nabla \times \mathbf{B})$ is the current density, and $\gamma$ is the adiabatic index.
By using the Els\"{a}sser variables \citep{1950PhRv...79..183E}, defined as:
\begin{equation}
\mathbf{z}^\pm = \mathbf{v} \pm \frac{\mathbf{B}}{\sqrt{\mu \rho}},
\label{Elsasser}
\end{equation} 
and considering only incompressible motions ($\nabla \cdot \mathbf{v} = 0$), the system of Eqs.~\ref{continuity}-\ref{divb} can be rewritten in the form \citep{1950PhRv...79..183E}:
\begin{align}
\label{eqzplus}
&\frac{\partial \mathbf{z}^+}{\partial t} + \mathbf{z}^- \cdot \nabla \mathbf{z}^+ = - \nabla P,\\
&\frac{\partial \mathbf{z}^-}{\partial t} + \mathbf{z}^+ \cdot \nabla \mathbf{z}^- = - \nabla P,\\
\label{eqzdiv}
&\nabla \cdot \mathbf{z}^\pm = 0,
\end{align}
where $P = p + \frac{\mathbf{B^2}}{2\mu}$ is the total pressure.
In the following, we consider an infinite and homogeneous medium, with a straight magnetic field $\mathbf{B}_0 = B_0 \mathbf{\hat{x}}$, where $\mathbf{\hat{x}}$ is the $x$-axis unit vector in Cartesian coordinates. This implies that the gradients of equilibrium quantities vanish. We consider perturbations of velocity and magnetic field of arbitrary magnitude over this equilibrium, such that $\mathbf{v} = (\mathbf{v}_0 = 0) + \mathbf{v}'$ and $\mathbf{B} = \mathbf{B}_0 + \mathbf{B}'$, where the zero subscript denotes the equilibrium values and the primed variables are perturbations. Then, we can rewrite the Els\"{a}sser variables in the form:
\begin{equation}
\mathbf{z}^\pm = \mathbf{z}_0^\pm + \mathbf{z}'^\pm = \pm v_{A0} \mathbf{\hat{x}}  + \left(  \mathbf{v}'  \pm \frac{\mathbf{B}'}{\sqrt{\mu \rho_0}} \right),
\label{Elspert}
\end{equation}
where $v_{A0} = \frac{B_0}{\sqrt{\mu \rho_0}}$ is the equilibrium Alfv\'en speed. Rewriting Eqs.~\ref{eqzplus}-\ref{eqzdiv} now yields:
\begin{align}
\label{eqzpertplus}
&\frac{\partial \mathbf{z}^+}{\partial t} + v_{A0} \frac{\partial \mathbf{z^+}}{\partial x} =   -\mathbf{z}^- \cdot \nabla \mathbf{z}^+ - \nabla P,\\
&\frac{\partial \mathbf{z}^-}{\partial t} - v_{A0} \frac{\partial \mathbf{z^-}}{\partial x} =   -\mathbf{z}^+ \cdot \nabla \mathbf{z}^- - \nabla P,\\
\label{eqzpertdiv}
&\nabla \cdot \mathbf{z}^\pm = 0,
\end{align}
where we dropped the prime from the perturbed Els\"{a}sser variables. Note that we did not linearize the system: perturbations can be of arbitrary amplitude. As noted in the Introduction, in the incompressible framework the Els\"{a}sser variables represent Alfv\'en waves propagating parallel or anti-parallel to the main magnetic field. This can be easily seen then by considering only one of the Els\"{a}sser variables nonzero in Eqs.~\ref{eqzpertplus}-\ref{eqzpertdiv}, i.e. either $\mathbf{z}^+ \ne 0, \mathbf{z}^- = 0$ or $\mathbf{z}^+ = 0, \mathbf{z}^- \ne 0$, which leads to two uncoupled equations:
\begin{align}
&\frac{\partial \mathbf{z}^+}{\partial t} = - v_{A0} \frac{\partial \mathbf{z^+}}{\partial x} &\mathbf{z}^- = 0, \\
&\frac{\partial \mathbf{z}^-}{\partial t} =  v_{A0} \frac{\partial \mathbf{z^-}}{\partial x} & \mathbf{z}^+ = 0,  
\end{align}

\begin{figure}[t]
  \centering
  \medskip
  \includegraphics[width=0.5\textwidth]{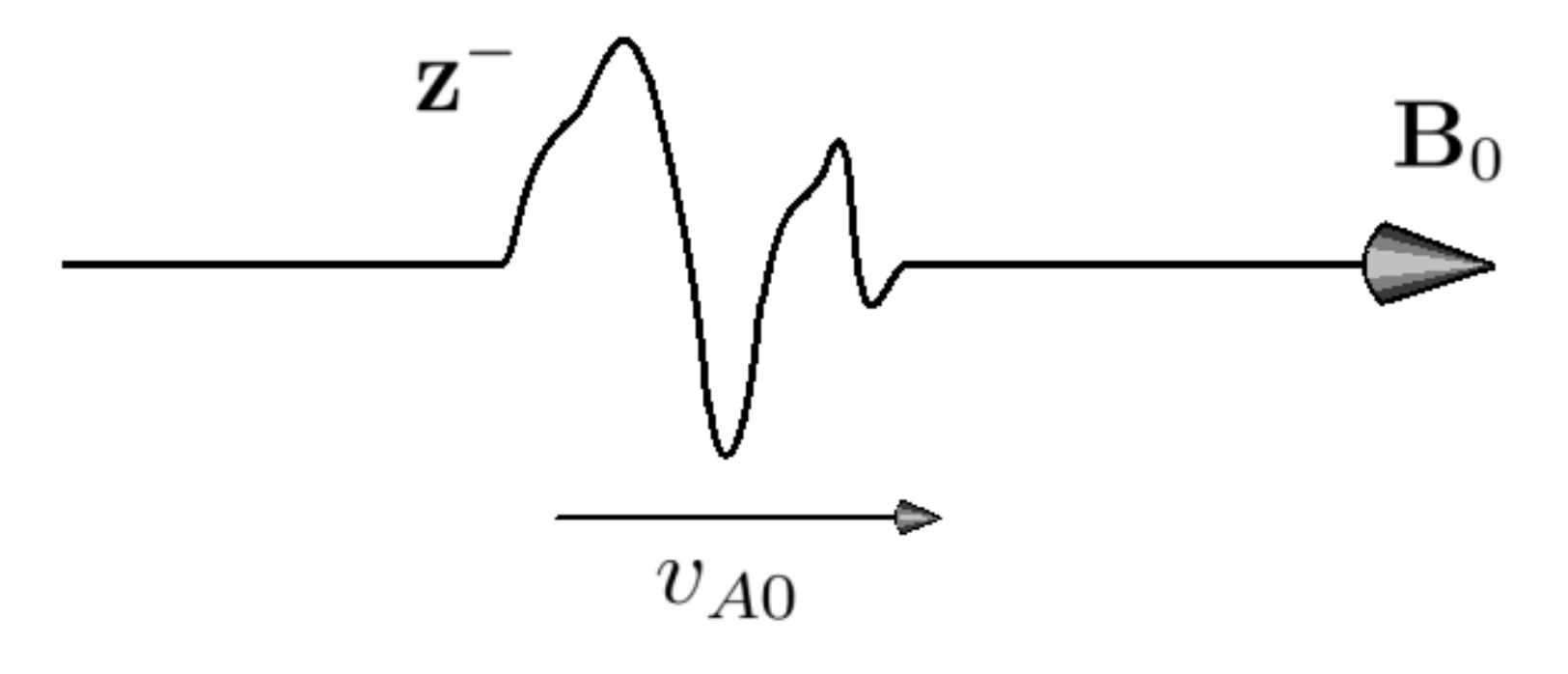}
  \caption{An Alfv\'en wave packet propagating parallel to the magnetic field, with speed $v_{A0}$}
  \label{picz}
\end{figure} 

while still $\nabla \cdot \mathbf{z}^\pm = 0$. Note that the total pressure gradient term is equal to zero if one of the Els\"{a}sser variables vanish. These equations describe arbitrary nonlinear pure Alfv\'en wave packages, propagating unidirectionally (see Fig.~\ref{picz}), with the exact solutions:
\begin{equation}
\mathbf{z}^\pm = \mathbf{z}(x \pm v_{A0} t).
\end{equation}
It is this property of the Els\"{a}sser variables that is usually exploited in turbulence studies. \par
Now, let us investigate the appearance of magnetoacoustic modes in the Els\"{a}sser formalism.  For this, first we return to the original velocity and magnetic field formulation in Eqs.~\ref{continuity}-\ref{divb}. Again we restrict ourselves to a uniform and homogeneous medium, over which we impose linear perturbations of all variables. 
 By differentiating the linearized form of Eq.~\ref{motion} with respect to time and substituting the time derivatives of the other variables in, and after some algebraic manipulation, a generalised wave equation for $\mathbf{v}'$ is obtained:
\begin{equation}
\label{wave_eq}
\frac{\partial^2 \mathbf{v}'}{\partial t^2} = c_s^2 \nabla (\nabla \cdot \mathbf{v}') + \{\nabla \times [\nabla \times (\mathbf{v}' \times \mathbf{B}_0)]\} \times \frac{\mathbf{B}_0}{\mu \rho_0},
\end{equation}
where 
\begin{equation}
c_s^2 = \frac{\gamma p_0}{\rho_0} = \gamma R_{sp} T_0
\end{equation}
is the square of the sound speed, with $R_{sp} = k_B/m$ the specific gas constant, $k_B$ the Boltzmann constant, $m$ the average mass per particle, and $T_0$ the equilibrium temperature. 
In the following, we will express the magnetic field in units for which $\mu = 1$. The wave equation (Eq.~\ref{wave_eq}) admits as solution waves which propagate vorticity and no compression (Alfv\'en waves) and waves which propagate compression but no vorticity (magnetoacoustic waves). 
\begin{figure}[t]
  \centering
  \medskip
  \includegraphics[width=0.5\textwidth]{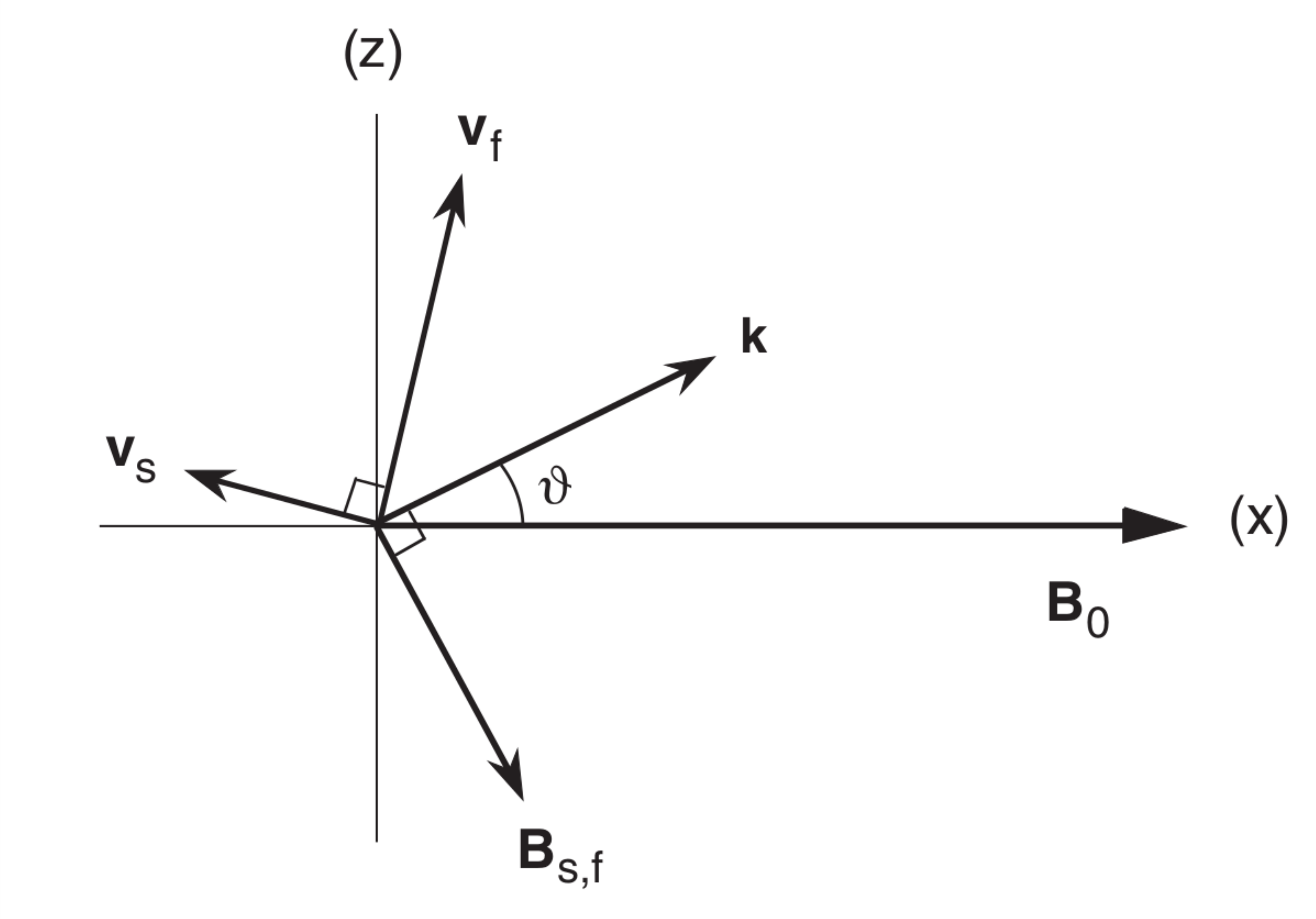}
  \caption{Velocity and magnetic field perturbations for magnetoacoustic waves. $\mathbf{k}$ is the wave vector, while $\nu$ denotes the angle between $\mathbf{B}_0$ and $\mathbf{k}$. Adapted from \citet{2004prma.book.....G}}.
  \label{eigensonic}
\end{figure} 
Magnetoacoustic waves are of two types: fast and slow. In a homogeneous and infinite medium, these three linear wave modes are uncoupled and have well-defined eigenfunctions. For the magnetoacoustic modes, the velocity, magnetic field, and density perturbations are \citep[][see also Fig.~\ref{eigensonic}]{2004prma.book.....G}:
\begin{align}
\mathbf{v}'_{s,f} &= A \left( \alpha_{s,f} \frac{k_\parallel}{k_\perp}, \quad 0,\qquad 1\qquad \right), \\
\mathbf{B}'_{s,f} &= \sqrt{\rho_0} A \left( \frac{v_{A0} k_\perp}{\omega_{s,f}},\quad0, \quad-\frac{v_{A0} k_\parallel}{\omega_{s,f}} \right), \\ 
\rho' &= \rho_0 A \frac{k_\perp}{\omega_{s,f}} \left( 1 + \frac{k^2_\parallel}{k^2_\perp} \alpha_{s,f} \right), 
\end{align}
where the subscript denotes slow or fast, $k_\parallel = \mathbf{k}\ \mathrm{cos}(\nu)$, $k_\perp = \mathbf{k}\ \mathrm{sin}(\nu)$, with $\nu$ the angle between $\mathbf{k}$ and $\mathbf{B}_0$ (see Fig.~\ref{eigensonic}), $\omega_{s,f}$ is the eigenfrequency of fast and slow waves \citep[for its expression see][]{2004prma.book.....G}, $A$ is the normalized velocity perturbation amplitude so that $|\mathbf{v}'_{s,f}| = 1$, and $\alpha_{s,f}$ is:
\begin{equation}
\alpha_{s,f} \equiv 1 - \frac{k^2 v_{A0}^2}{\omega_{s,f}^2}, \qquad \alpha_s \leq 0\  \mathrm{and}\ \alpha_f \geq 0.
\end{equation} 
Note that, due to the presence of density perturbations, the perturbed Els\"{a}sser fields are expressed now as: 
\begin{equation}
\mathbf{z}^\pm = \mathbf{z}_0^\pm + \mathbf{z}'^\pm = \mathbf{v}_0 + \mathbf{v}' \pm \frac{\mathbf{B}_0 + \mathbf{B}'}{\sqrt{\rho_0 + \rho'}} = \mathbf{v}_0 \pm v_{A0} \mathbf{\hat{x}} + \left( \mathbf{v}' \pm \frac{\mathbf{B}'}{\sqrt{\rho_0}}  \mp \frac{\rho'}{2 \rho_0}v_{A0}\mathbf{\hat{x}} \right),
\label{Elspertcomp}
\end{equation} 
where the difference from its incompressible counterpart in Eq.~\ref{Elspert} is the presence of an additional term along $\mathbf{B}_0$. Now, by plugging in the eigenfunctions for magnetoacoustic waves in the above expression, we obtain the perturbed Els\"{a}sser fields for fast and slow waves:
\begin{equation}
\mathbf{z}^\pm_{s,f} = A \left(\alpha_{s,f} \frac{k_\parallel}{k_\perp} \pm \frac{k_\perp v_{A0}}{2 \omega_{s,f}} \left( 1 - \frac{k^2_\parallel}{k^2_\perp} \alpha_{s,f} \right),\qquad 0, \qquad1 \mp \frac{k_\parallel v_{A0}}{\omega_{s,f}}\right).
\label{elsassersonic}
\end{equation}
\begin{figure}[t]
  \centering
  \medskip
  \includegraphics[width=0.4\textwidth]{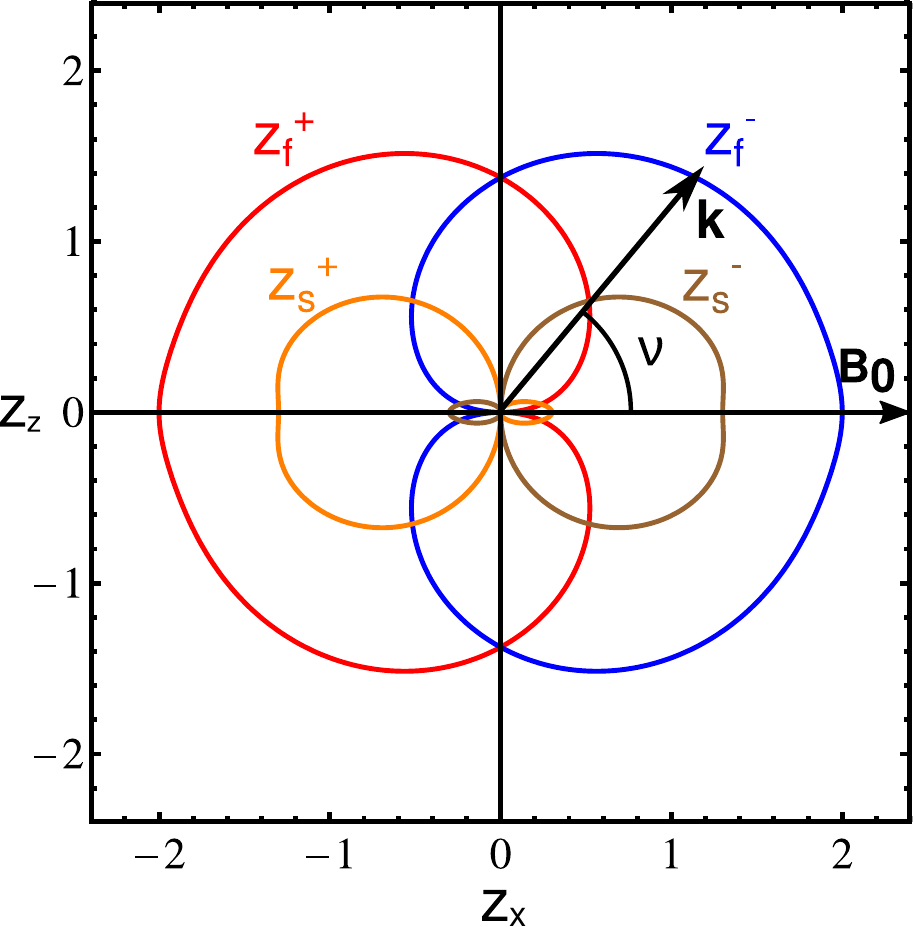}
  \caption{Polar plot of Eq.~\ref{elsassersonic}, representing the magnitude of $\mathbf{z}^\pm_{s,f}$ normalized by multiplying with the phase speed $\omega_{s,f}/|k|$, as a function of the angle $\nu$ between $\mathbf{k}$ and $B_0 \mathbf{\hat{x}}$. The parameters used are: $v_{A0} =1,c_s = 0.8,\rho_0 = 1,k = 1$.} 
  \label{elsasserphase}
\end{figure} 
By looking at the diagram of Eq.~\ref{elsassersonic} plotted in Fig.~\ref{elsasserphase}, we can appreciate that magnetosonic waves are described by both $\mathbf{z}^+$ and $\mathbf{z}^-$, i.e. a single magnetosonic wave presents perturbations in both variables while propagating. Therefore, these waves cannot be separated in `inward' or `outward' propagating waves with respect to the background magnetic field by using the Els\"{a}sser variables. As exception, for parallel propagation i.e. $\mathbf{k} \parallel \mathbf{B}_0$, fast waves are described by only one of the Els\"{a}sser variables, the selection depending on the propagation direction. Slow waves present both variables for strictly parallel propagation, albeit with different amplitudes. For perpendicular propagation, fast waves are described by both Els\"{a}sser variables, with equal magnitude. \par 
We reiterate that the analysis above is only valid for a homogeneous and infinite medium. As mentioned earlier, when inhomogeneities are present, in general waves cannot be separated (i.e. they do not possess separate eigenfunctions and frequencies as shown in the analysis above), as they are linearly coupled. Without entering the complicated mathematical treatment of MHD waves in an inhomogeneous plasma, this generally means that a single wave has both Alfv\'en and magnetoacoustic properties, i.e. mixed properties: it propagates both vorticity and compression, is driven by both magnetic tension and pressure, etc. \citep{2011SSRv..158..289G}. Translated into the 
Els\"{a}sser picture this implies that due to the partial magnetoacoustic character of waves with mixed properties, they are generally described by perturbations in both Els\"{a}sser variables, both propagating in the direction of the wave vector $\mathbf{k}$. This property of waves in inhomogeneous media will be presented in Section~\ref{four}.

\section{Simulation of MHD waves in a 2.5D model} \label{three}

In order to demonstrate and help visualizing the results derived in the previous section, we run ideal 2.5D MHD simulations using the code \texttt{MPI-AMRVAC} \citep{2012JCoPh.231..718K,2014ApJS..214....4P,2018ApJS..234...30X}. Here 2.5D means 2 spatial dimensions and 3 vector components, i.e. perturbations along the third direction are supposed to have zero wavenumber along that direction. We use the implemented one-step \texttt{tvd} method with Roe's solver and \texttt{Woodward} slope limiter. The constraint on the magnetic field divergence is maintained using Powell’s scheme. On the square numerical domain of size $(-L/2,L/2)^2$, the uniform resolution is $384^2$ cells. Convergence studies with double the resolution show no important differences in the dynamics. We use open boundary conditions, however, the simulation stops before the wave-fronts reach the boundaries.\par
\begin{figure}[t]
  \centering
  \medskip
  \includegraphics[width=1.0\textwidth]{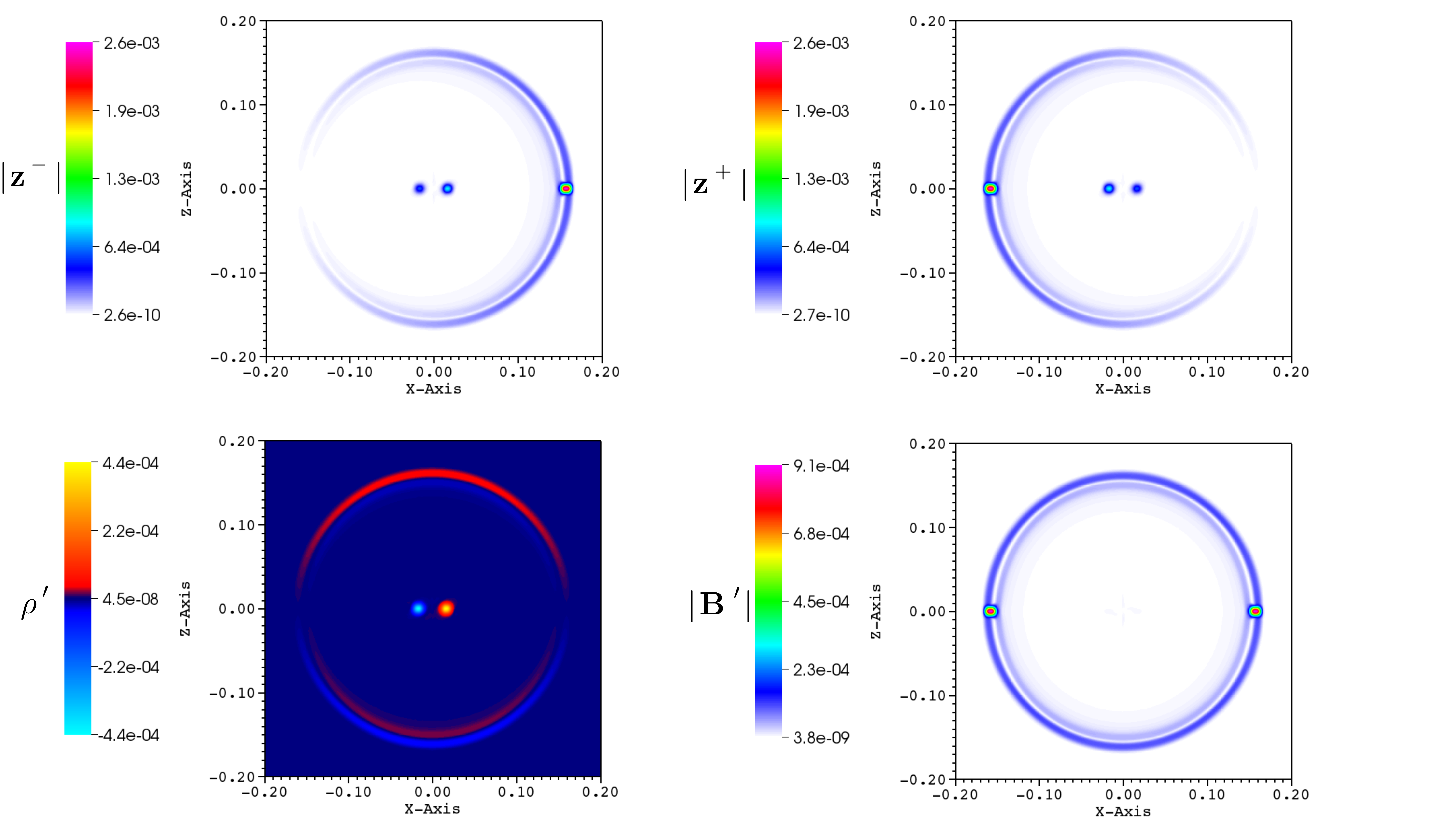}
  \caption{Snapshots from the 2.5D simulation, showing $|\mathbf{z}^-|$ (\textit{top-left}), $|\mathbf{z}^+|$ (\textit{top-right}), $\rho'$ (\textit{bottom-left}), and $|\mathbf{B}'|$ (\textit{bottom-right}), at some time $t_f$ before the waves reach the boundaries. Plot and axis values are in user units. (In the online version of the paper, the snapshots are animated, representing their evolution from $t_0 = 0$ to $t_f$).} 
  \label{2.5D}
\end{figure}
\begin{figure*}[h]
    \centering
       \begin{tabular}{@{}cc@{}}
        \includegraphics[width=0.34\textwidth]{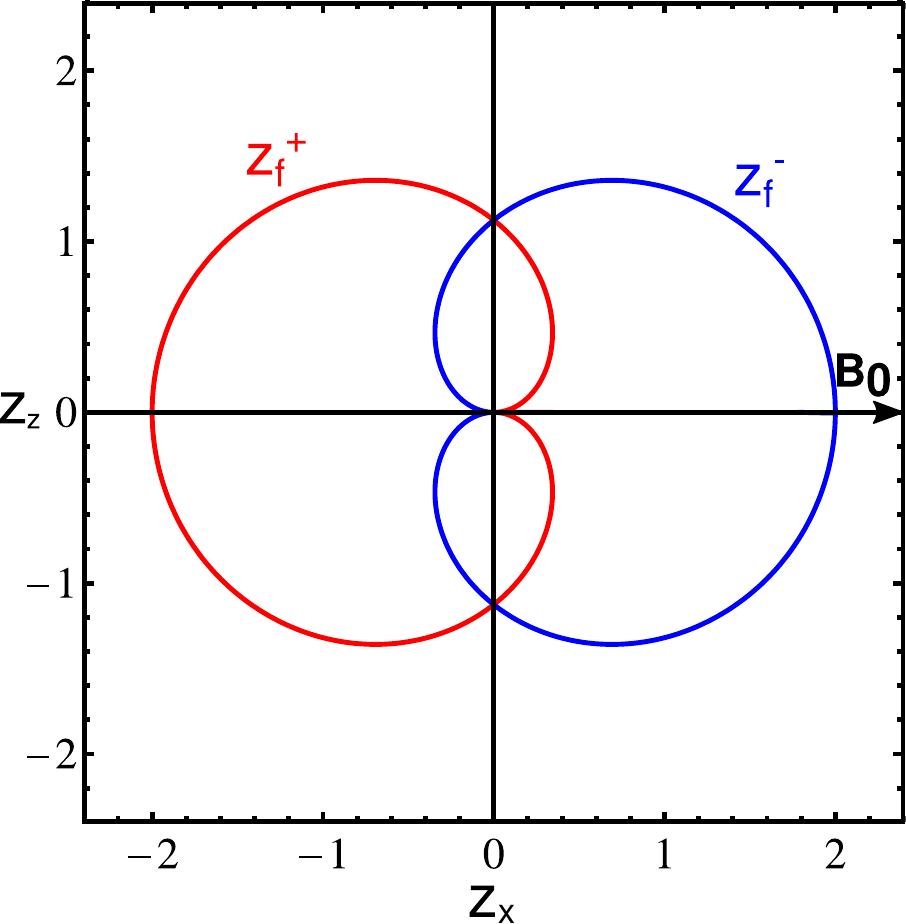}  
        \includegraphics[width=0.36\textwidth]{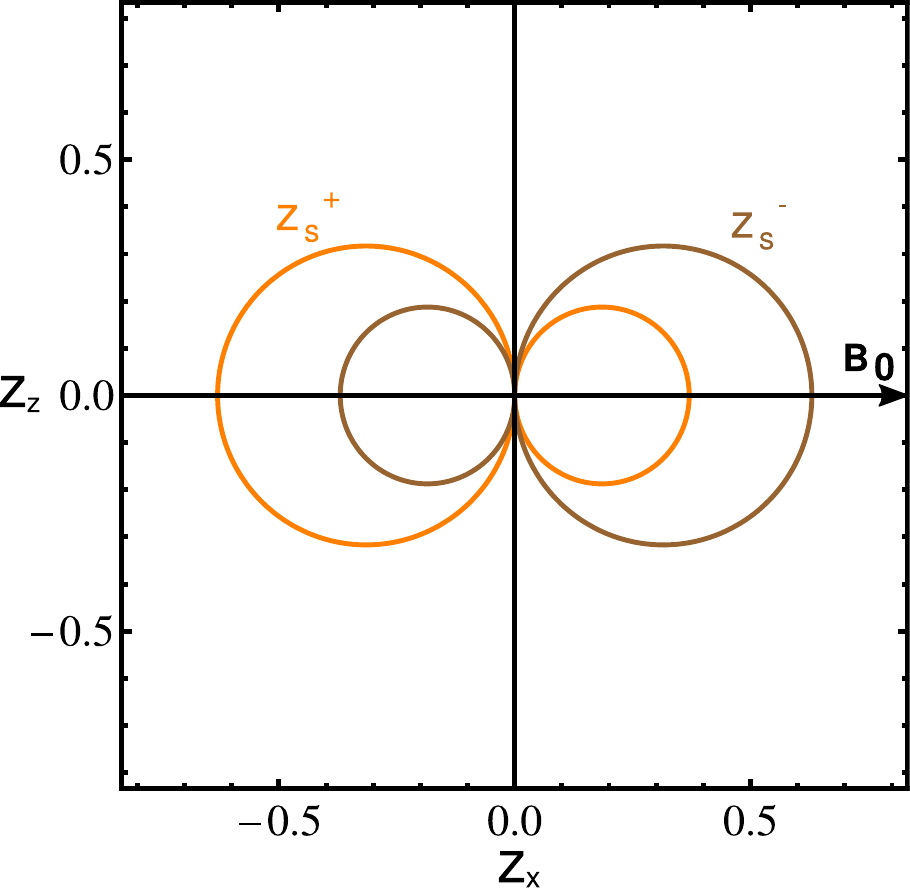} \\
       \end{tabular}  
        \caption{Same as in Fig.~\ref{elsasserphase}, but for parameters in accordance with the 2.5D simulation parameters. \textit{Left:} Fast waves. \textit{Right:} Slow waves.}
        \label{2.5D_anal}
 \end{figure*}

The equilibrium consists of a homogeneous plasma without flows, with a straight, homogeneous magnetic field $\mathbf{B}_0 = B_0 \mathbf{\hat{x}}$, with plasma $\beta = c_s^2/V_{A0}^2 \approx 0.013$. We trigger linear MHD waves by considering an initial pulse in all 3 components of velocity:
\begin{equation}
v'_{x,y,z}(t=0) = M\ \mathrm{exp} \left( - \frac{x^2 + z^2}{R^2} \right),
\label{2.5Dpert}
\end{equation}
where $M = 2 \cdot 10^{-3}$ is the Alfv\'en Mach number for the $y,z$ components, and the sonic Mach number for the $x$ component, and $R \approx 0.034\ L$ is the pulse width. We use these small Mach numbers in order to minimize nonlinear couplings between the fast, slow, and Alfv\'en waves. Note that as the sound speed is smaller than the Alfv\'en speed, the perturbation amplitude of $v'_x$ is correspondingly smaller. This is done in order to have the same degree of nonlinearity for all waves. The resulting wave behaviour is shown in Fig.~\ref{2.5D}.
By a closer inspection of the top graphs in Fig.~\ref{2.5D}, one can distinguish the specific appearance of fast, slow, and Alfv\'en waves as expressed through the Els\"{a}sser variables. As described in Section~\ref{two}, pure Alfv\'en waves are necessarily described by only one of the Els\"{a}sser variables: this can be seen as the strong pulse propagating along $\mathbf{B}_0$ in the top-left snapshot ($\mathbf{z}^-$), and to the left in the top-right snapshot ($\mathbf{z}^+$). Fast and slow waves, on the other hand, are described by both ($\mathbf{z}^-$) and ($\mathbf{z}^+$), as expressed in Eq.~\ref{elsassersonic}. Note that while the slow waves present both Els\"{a}sser variables for propagation along the magnetic field, the fast waves share the property of Alfv\'en waves when propagating parallel to the background magnetic field. For a comparison with the analytical results, see Fig.~\ref{2.5D_anal}.

\section{Simulation of linearly coupled MHD waves in a 3D inhomogeneous model} \label{four}

\begin{figure}[t]
  \centering
  \medskip
  \includegraphics[width=0.5\textwidth]{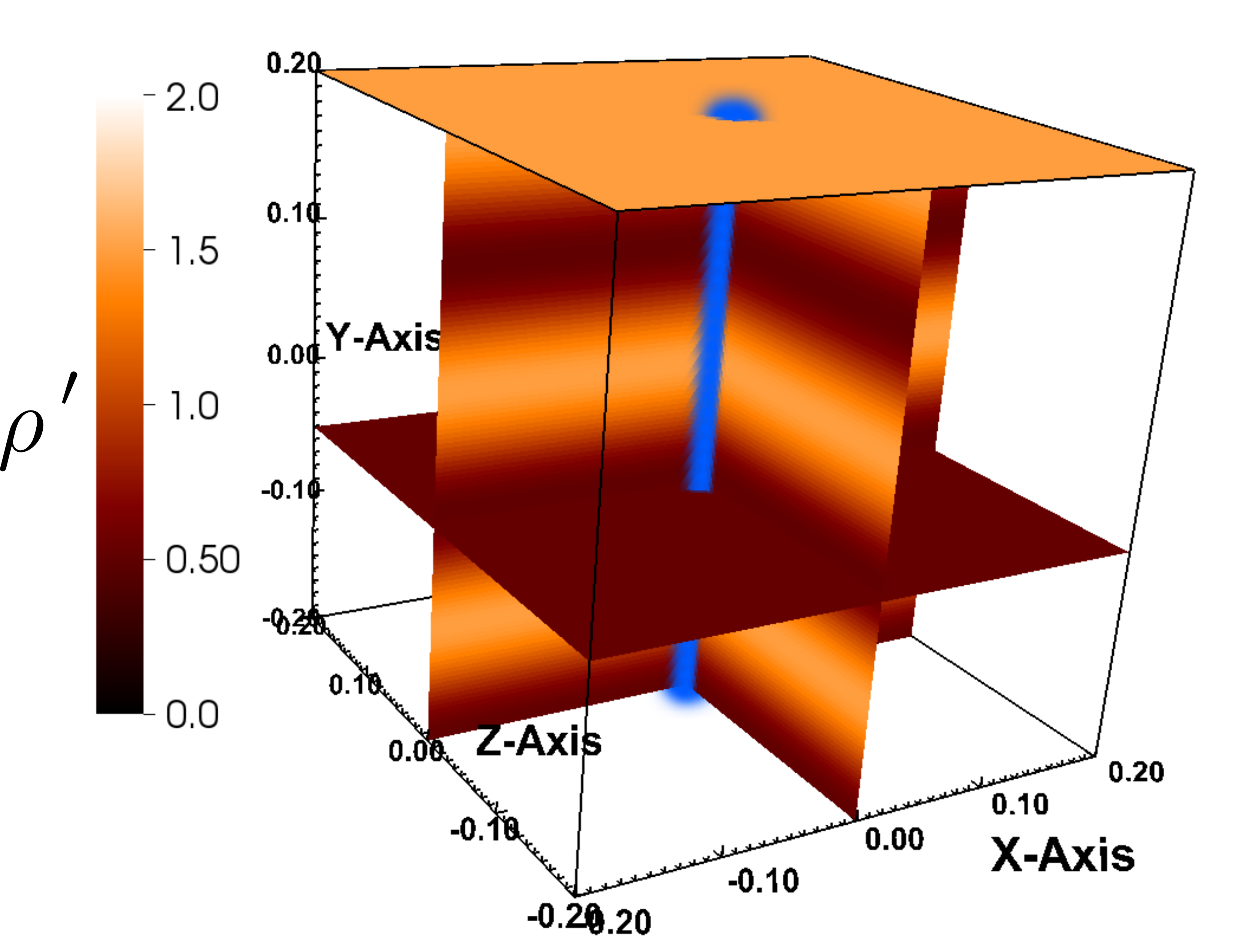}
  \caption{Multi-slice snapshot of the initial condition for the 3D simulation, showing density (with the associated color scale, in user units) and the initial perturbation in $\mathbf{v}'_y$ (`cylinder' around $x$ = 0, $z = 0$). } 
  \label{3Dsetup}
\end{figure}

\begin{figure}[t]
  \centering
  \medskip
  \includegraphics[width=1.0\textwidth]{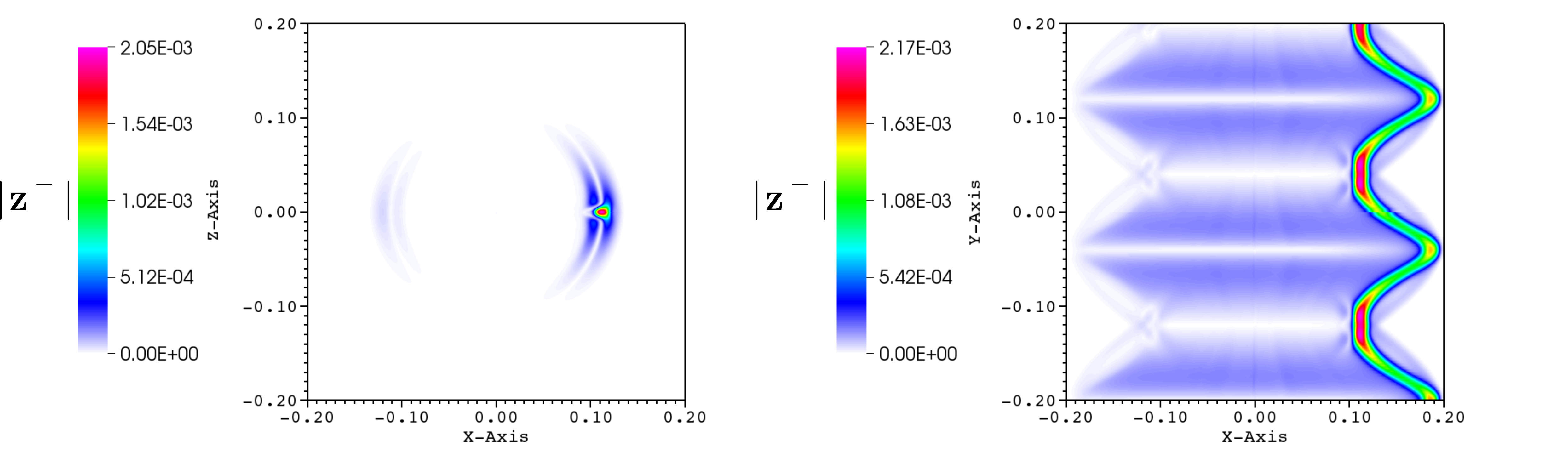}
  \caption{Slice snapshots from the 3D simulation, showing $|\mathbf{z}^-|$ in the $x-z$ plane at $y = 0.04$ (\textit{left)}, and in the $x-y$ plane at $z=0$ (\textit{right}), at some time $t_f$ before the waves reach the boundaries. Plot and axis values are in user units. $|\mathbf{z}^+|$ is not shown for brevity, as it is the mirror image of $|\mathbf{z}^-|$ with respect to the $x=0$ plane. (In the online version of the paper, the snapshots are animated, representing their evolution from $t_0 = 0$ to $t_f$).} 
  \label{3D}
\end{figure}

In the previous section, by employing a 2.5D model, i.e. considering no variation along the $y$-axis, we achieved the linear decoupling of the fast, slow and Alfv\'en waves \citep{1998A&A...335..329D,2000A&A...356..724D,2011SSRv..158..289G}. The 3D model employed in this section can be viewed as an extension of the previous 2.5D model in the $y$-direction. The code, numerical methods, and solvers used are the same as in the previous section. The resolution is $256^2 \times 128$, with less resolution in the slow-varying $y$-direction. The cubic domain size is $(-L/2,L/2)^3$. Again we conducted convergence studies, and found that there are essentially no differences compared to higher resolution runs.  In order for linear coupling to occur, we consider density variations along the $y$-direction. In this equilibrium, waves cannot be separated into pure fast, slow, and Alfv\'en waves. The density variation is described by:
\begin{equation}
\rho(y) = \rho_0 + \frac{1}{2}\rho_0\ \mathrm{sin} \left( \frac{5 \pi}{L} y \right),
\end{equation}
where $L$ and $\rho_0$ is the same as for the 2.5D simulation. In order to show that MHD waves in this equilibrium are indeed linearly coupled, we only perturb the $y$-component of the velocity initially, which necessarily leads to perturbations in the other components once the simulations starts. The perturbation for $v'_y$ is the same as for the 2.5D simulation (Eq.~\ref{2.5Dpert}). Note that the perturbation does not depend on $y$, i.e. it acts along the entire $y$-direction with the same magnitude (see Fig.~\ref{3Dsetup}). 
We run the simulation until $t_f$, coinciding with the time when the first wave-front reaches the lateral boundary. The evolution of the Els\"{a}sser variables can be seen in Fig.~\ref{3D}. 
Note the presence of a weaker anti-parallel component (towards negative $x$-axis values) of $\mathbf{z}^-$: the appearance of this component is the result of linear coupling of MHD waves in the inhomogeneous plasma, as explained in Section~\ref{two}.  The ratio of amplitudes of the left and right propagating $\mathbf{z}^-$ is 0.03, while the peak density perturbation is $\approx 10^{-5} \rho_0$ in this case. We have also measured quantities which reflect the Alfv\'enic and compressive component of the fluctuations \citep[see, e.g.][]{2016JPlPh..82f5302C}, namely the normalized total magnetic field perturbation ($|\delta \mathbf{B}|/|\mathbf{B}_0| \approx 0.0117$) and the perturbation of the magnetic field magnitude ($\delta |\mathbf{B}|/|\mathbf{B}_0| \approx 0.000134$), respectively. Note that the ratio of compressive to Alfv\'enic fluctuations is around $1\%$, which is usually the ratio found within the fast solar wind \citep{1971JGR....76.3534B,2013LRSP...10....2B}. These values reflect the highly Alfv\'enic nature of the perturbation. In the cut along the $y$-direction, the apparent phase mixing of the waves can be seen. Phase mixing occurs due to the presence of a variable Alfv\'en speed profile, and results in the curved appearance of the wave-fronts (see right panel of Fig.~\ref{3D}).  Note however that this `phase mixing' is different than the one described in \citet{1983A&A...117..220H}, as in this case the density variation is in the direction of the perturbation \citep[see][]{1991ApJ...376..355P}. Furthermore, the variable amplitude of $\mathbf{z}^-$ along the $y$-direction is evident: the amplitude varies approximately with the local equilibrium density gradient. Also, we can observe waves propagating faster than the `main' phase mixed $v'_y$ wave-front, seen as `bulges' atop these, when the propagation direction is oblique to the magnetic field direction. This can also be seen in the anti-parallel component, leading to the wave-front `crosses' seen along $x \approx -0.1$ in the $x-y$ slice. The $x-z$ slice in Fig.~\ref{3D} is at the $y$-axis position coinciding with the position of one of these crosses. In the $x-y$ slice, a continuous presence of $\mathbf{z}^-$ spanning the $x$-axis from the parallel to the anti-parallel propagating wave fronts can be seen. This is due to the density perturbations present near the $z=0$ plane, which represent non-propagating entropy or thermal modes. The amplitude variation of the entropy mode along the $y$-axis approximately follows the local density gradient. This  linear coupling of the entropy mode and propagating modes in the presence of inhomogeneities will be investigated in more detail in another study. These perturbations then manifest in the Els\"{a}sser fields through the additional term in Eq.~\ref{Elspertcomp}. In Fig.~\ref{3D} the $y$-axis position of the $x-z$ slice was also chosen in order to exclude these non-propagating modes. \par
\begin{figure}[t]
  \centering
  \medskip
  \includegraphics[width=1.0\textwidth]{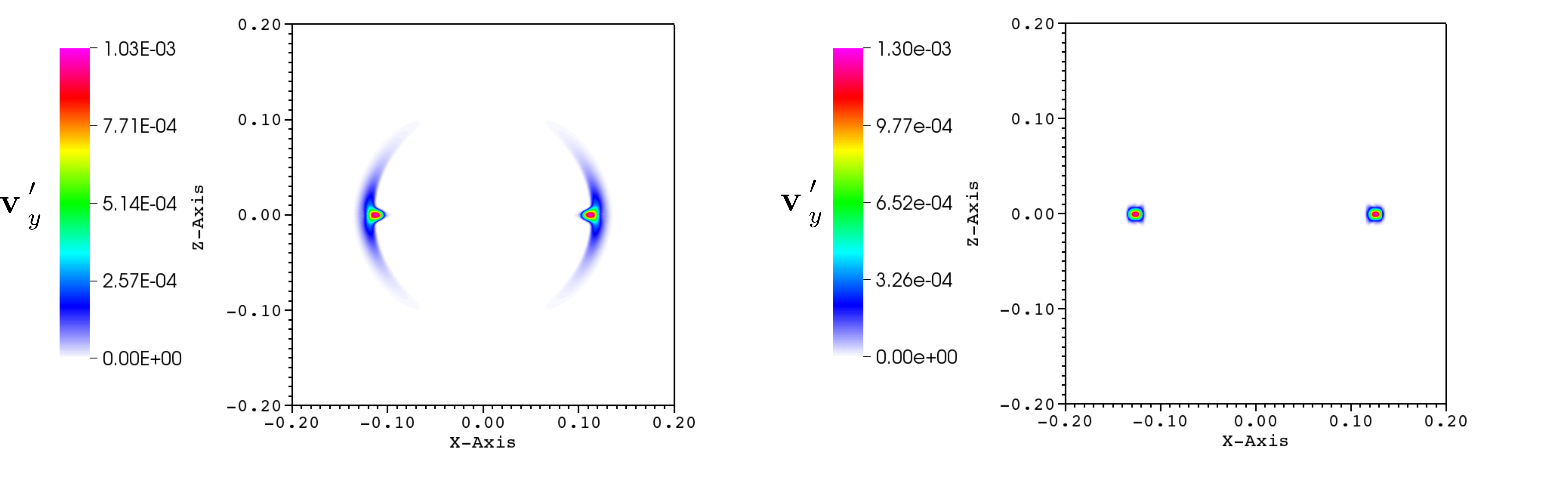}
  \caption{Slice snapshots from the inhomogeneous 3D simulation (\textit{left}) and the homogeneous 3D simulation (\textit{right}) , showing $\mathbf{v}'_y$ in the $x-x$ plane at $y=0.04$, at some time $t_f$ before the waves reach the boundaries. Plot and axis values are in user units. (In the online version of the paper, the snapshots are animated, representing their evolution from $t_0 = 0$ to $t_f$).} 
  \label{v3}
\end{figure}

In order to see how the linearly coupled evolution differs from the homogeneous evolution, we ran a 3D simulation identical to the one described at the beginning of this section, except we do not consider density variations along the $y$-axis. Therefore, we set the density everywhere to $\rho_0$. In this setting, the initial perturbation corresponds to superposed pure Alfv\'en waves which, once the simulations starts, separate into parallel and anti-parallel propagating pure Alfv\'en pulses. A comparison with the inhomogeneous evolution of $v'_y$ can be seen in Fig.~\ref{v3}. 
The obvious difference is the smeared appearance of the $v'_y$ component in the linearly coupled (inhomogeneous) case. Indeed, this can be interpreted as the manifestation of a wave with mixed Alfv\'en-fast properties. In the homogeneous, or the 2.5D case, the only perturbation along the $y$-direction is due to the pure Alfv\'en wave, which retains its shape as it propagates away from the origin.

\section{Conclusion}\label{five}

The Els\"{a}sser formalism is a very useful approach to incompressible MHD, since it transforms the usual velocity-magnetic field picture into a more intuitive symmetric system of equations. This symmetric system is interpreted as the interaction of pure Alfv\'en waves propagating parallel and anti-parallel to the main magnetic field, each completely described by one of the Els\"{a}sser variables. Based on this fact, numerous previous studies in MHD turbulence employed these variables to separate between parallel and anti-parallel propagating waves, even in inhomogeneous and compressible plasmas, such as the solar wind. While this separation is strictly valid in a homogeneous and incompressible plasma, we show that once we account for the presence of compressibility and inhomogeneities, the Els\"{a}sser variables cannot be used anymore to fully separate wave modes propagating in opposite directions. Even under homogeneous conditions, when the waves are linearly decoupled, magnetoacoustic waves, i.e. fast and slow waves are necessarily described by both Els\"{a}sser variables, propagating in the same direction, i.e. in the direction of wave vector $\mathbf{k}$. Once plasma inhomogeneities are present, waves cease to exist in their pure form, and we can no longer classify the waves as being Alfv\'en, fast or slow. The linear coupling of waves due to plasma inhomogeneity means that waves have in general mixed properties. We show using 3D inhomogeneous simulations that in this case, an initial pulse perpendicular to the background magnetic field results in waves which propagate both Els\"{a}sser fields.  Nevertheless, even under inhomogeneous conditions, in situations in which the waves are highly Alfv\'enic (dominantly having the properties of a pure Alfv\'en wave), the Els\"{a}sser fields appear to be able to indicate the dominating propagating direction to a very good approximation, at least in the present study. That is, the pulse propagating along the magnetic field shows a clearly dominant $\mathbf{z}^-$ ($\approx 97\%$) and a weaker $\mathbf{z}^+$ ($\approx 3\%$) component, and the pulse propagating in the opposite direction the other way around.  However, in an earlier study simulating driven, unidirectionally propagating Alfv\'enic waves, we found that the amplitude ratio between the dominant and weaker Els\"{a}sser fields was on average 0.1 \citep{2017NatSR...14820M}, while the ratio of compressive to Alfv\'enic fluctuations was also on the order of $1-2\%$. Furthermore, at points the `anomalous' Els\"{a}sser field could surpass the principal component. Therefore, at the present moment we are unable to determine the precision of the Els\"{a}sser formalism to separate inward and outward propagating waves under inhomogeneous and compressible conditions, as it might depend on many factors. We suggest that two of these factors determining the relative amplitude of the two Els\"{a}sser fields are the local density gradient (degree of inhomogeneity) and the wavenumber across and along the magnetic field. However, the detailed analysis concerning the ratio of Els\"{a}sser field amplitudes varying as a function of these factors is aimed as a follow-up study. This paper was rather intended as a first demonstration of the fact that mangetoacoustic and linearly coupled MHD waves are generally described by both Els\"{a}sser variables and therefore cannot be strictly separated in parallel and anti-parallel propagating components.
                              
\begin{acknowledgements} T.V.D. was supported by the GOA-2015-014 (KU Leuven) and the European Research Council (ERC) under the European Union's Horizon 2020 research and innovation programme (grant agreement No. 724326). \end{acknowledgements}

\bibliographystyle{apj} 
\bibliography{../Biblio}{} 

\end{document}